\documentclass[aps,amsmath,amssymb,eqsecnum,preprint,showpacs]{revtex4}
\usepackage{amsfonts}

\def \D {\mbox{D}}

\begin{document}
\title{Bianchi IX Brane-world Cosmologies}

\author                 {R. J.  van den Hoogen}
\email                  {rvandenh@stfx.ca}
\affiliation            {Department of Mathematics, Statistics, and Computer
Science,
                        Saint Francis Xavier University, Antigonish, N.S.,
                        B2G 2W5, Canada}

\author                 {A. A. Coley}
\email                  {aac@mscs.dal.ca}
\affiliation            {Department of Mathematics and Statistics,
                        Dalhousie University, Halifax, N.S.,B3H 3J5, Canada}

\author                 {Y. He}
\email                  {yanjing@mscs.dal.ca}
\affiliation            {Department of Mathematics and Statistics,
                        Dalhousie University, Halifax, N.S.,B3H 3J5, Canada}

\begin{abstract}
We shall investigate the asymptotic properties of the Bianchi type IX cosmological model in the Brane-world scenario.  The matter content is assumed to be a combination of a perfect fluid and a minimimally coupled scalar field that is restricted to the Brane.  A detailed qualitative analysis of the Bianchi type IX braneworld containing a scalar field having an exponential potential is undertaken. It is found that the Brane-Robertson Walker solution is a local source for the expanding Bianchi IX models, and if $k^2<2$ the expanding Bianchi IX models asymptote to the power-law inflationary solution.  The only other local sink is the contracting Brane-Robertson Walker solution.  An analysis of the Bianchi IX models with a scalar field with a general potential is discussed, and it is shown that in the case of expanding models, for physical scalar field potentials close to the initial singularity, the scalar field is effectively massless, and the solution is approximated by the Brane-Robertson Walker model. 
\end{abstract}


\maketitle{}


\section{Introduction}

It is believed that Einstein's General Relativity breaks down at sufficiently high energies. Developments in string theory suggest that gravity may truly be a higher dimensional theory, becoming an effective $4$-dimensional theory at lower energies.  Some researchers have suggested an alternative scenario in which the matter fields are restricted to a $3$-dimensional Brane-world embedded in $1+3+d$ dimensions (the bulk), while the gravitational field is free to propagate in the d extra dimensions \cite{rubakov}. In this paradigm it is not necessary for the $d$ extra dimensions to be small, or even
compact, a radical departure from the standard Kaluza-Klein scenario. Randall and Sundrum~\cite{randall} have shown that for $d=1$, gravity can be localized on a single $3$-brane even when the fifth dimension is infinite.   It has now become very important to test the astrophysical and cosmological implications of these higher dimensional theories derived from string theory.  Can these cosmological models derived from string theory explain the high degree of homogeneity and isotropy we currently observe?

The dynamical
equations on the 3-brane \cite{Maartens:2001b,Maartens:2000a,Shiromizu:2000} differ from the general relativity equations by terms that carry the effects of imbedding and of the
free gravitational field in the five-dimensional bulk. 
The local (quadratic) energy-momentum corrections are
significant only at very high energies and the dynamical equations reduce to the regular Einstein field equations of General Relativity for late times.  However, for very high energies (i.e, early times), these additional energy momentum correction terms will play a very critical role in the evolutionary dynamics of these Brane-world models.  In addition to  the matter field corrections, there are nonlocal effects (that modify the Friedmann equation on the brane in these models) from the free gravitational field in the bulk, transmitted via the projection ${\cal E}_{\mu\nu}$ of the bulk Weyl tensor, that contribute further corrections to the Einstein equations on the brane.

Cosmological observations indicate that we live in a universe which is
remarkably uniform on very large scales. However, the spatial homogeneity and
isotropy of the universe is difficult to explain within the
standard general relativistic framework since, in the presence of matter,
the class of solutions to the Einstein equations which evolve
towards a FRW universe is essentially a set of measure
zero~\cite{CollinsHawking:1973}. In the
inflationary scenario, we live in
 an isotropic region of a potentially highly 
irregular universe as the result of an expansion phase in the early universe
thereby solving many of the problems of cosmology. Thus this
scenario can successfully generate a homogeneous and
isotropic FRW-like universe from initial conditions which, in the
absence of inflation, would have resulted in a universe far
removed from the one we live in today. However, still only a restricted set
of initial data will lead to smooth enough conditions for the
onset of inflation (i.e., the so-called cosmic no-hair theorems
only apply to non-generic models -- see~\cite{js86,rellis86}), so the
issue of homogenization and isotropization is still not satisfactorily solved.
Indeed, the initial conditions problem, that is to explain why  the universe is so isotropic and spatial homogeneous from generic initial conditions,  is perhaps one of the central problems of modern theoretical cosmology.
These issues were recently revisited in the context of brane cosmology \cite{Coley:2002a,Coley:2002b}, and this is one of the motivations for the present work.

Indeed, researchers have investigated both anisotropic and isotropic Brane-world models, 
trying to ascertain whether the effects of the bulk gravitational field would
allow one to solve the isotropy problem.   A lot of effort has been directed at the so-called Friedmann Brane-world models \cite{cline,BinDefLan:2000}.  Dynamics of a brane-world universe filled with a perfect fluid have been intensively investigated during the last three years \cite{CamposSopuerta:2001a,CamposSopuerta:2001b,Coley:2002a,Coley:2002b}.  It has been found there exists new regimes that are not inherent in the standard cosmology, such as stable oscillation \cite{CamposSopuerta:2001b} and the collapse of a flat universe \cite{Santos:2001}. Some features of brane-world inflation have been studied in \cite{Maartens:2000b,Maartens:2001b,Copeland:2000} and the cosmological dynamics for exponential scalar field potentials have been described in \cite{Maeda:2002,Dunsby:2002}.

Shiromizo et al. \cite{Shiromizu:2000} have developed an elegant covariant approach to the bulk effects on the brane.  The equations derived by Maartens \cite{Maartens:2000a} are an extension of earlier work by Ellis and MacCallum that has been subsequently developed more recently in the book by Wainwright and Ellis \cite{Elliswainwright}.   Using the formalism developed by Maartens and Wainwright and Ellis, we propose to investigate the dynamical behavior in a wider class of anisotropic models than what has been previously analyzed.  Spatially homogeneous models of Class A (and in particular Bianchi type IX), containing both a perfect fluid and a scalar field will be investigated.  The resulting field equations will yield a system of ordinary differential equations, suitable for a geometric analysis using dynamical systems techniques.   This analysis will determine whether the dynamics of the Brane-world scenario mimics the dynamics of a General Relativistic cosmology at late times.  We are particularly interested in both the early-time (nature of the initial singularity) and late-time behaviour (i.e., whether these models inflate and isotropize).


\section{Governing Equations}

The field equations induced on the brane are derived via an
elegant geometric approach by Shiromizu et al. \cite{Shiromizu:2000,Maartens:2000a}, using
the Gauss-Codazzi equations, matching conditions and $Z_2$
symmetry. The result is a modification of the standard Einstein
equations, with the new terms carrying bulk effects onto the
brane:
\begin{equation}
G_{\mu\nu}=-\Lambda g_{\mu\nu}+\kappa^2
T_{\mu\nu}+\widetilde{\kappa}^4S_{\mu\nu} - {\cal E}_{\mu\nu}\,,
\end{equation}
where 
\begin{equation}
\kappa^2=\frac{8\pi}{M_{\rm p}^2}\,,~~\lambda=6{\frac{\kappa^2}{\widetilde\kappa^4}} \,, ~~ \Lambda =
\frac{4\pi}{ \widetilde{M}_{\rm p}^3}\left[\widetilde{\Lambda}+
\left({\frac{4\pi}{3\widetilde{M}_{\rm
p}^{\,3}}}\right)\lambda^2\right]\,.
\end{equation}
It is common to assume through fine tuning (a la Randall Sundrum) that the effective cosmological constant on the brane is zero, i.e., $\Lambda=0$, However,
we shall assume that it is non-zero but positive.

The brane energy-momentum
tensor for a perfect fluid and a minimally-coupled scalar field is given by
\begin{equation}
T_{\mu\nu}=T_{\mu\nu}^{\ \ perfect \ fluid}+T_{\mu\nu}^{\ \ scalar\ field},\label{source}
\end{equation}
where
\begin{eqnarray}
T_{\mu\nu}^{\ \ perfect \ fluid}&=&\rho u_\mu u_\nu+ph_{\mu\nu}\,\\
T_{\mu\nu}^{\ \ scalar\ field} &=& \phi_{;\mu}\phi_{;\nu}
-g_{\mu\nu}\left(\frac{1}{2}\phi_{;\alpha}\phi^{;\alpha}+V(\phi)\right),
\end{eqnarray}
where $u^\mu$ is the fluid 4-velocity, $\rho$ and $p$ are the energy density and isotropic
pressure, $\phi$ is the minimally coupled scalar field having potential $V(\phi)$, and the projection tensor $h_{\mu\nu} \equiv g_{\mu\nu}+u_\mu u_\nu$ projects orthogonal to $u^\mu$.   If $\phi_{;\mu}$ is timelike, then a scalar field with potential $V(\phi)$ is equivalent to a perfect fluid having an energy density and pressure 
\begin{eqnarray}
\rho^{scalar\ field}&=&-\frac{1}{2}\phi_{;\mu}\phi^{;\mu}+V(\phi)\\
p^{scalar\ field} &=&-\frac{1}{2}\phi_{;\mu}\phi^{;\mu}-V(\phi).
\end{eqnarray}

The bulk corrections to the Einstein equations on the brane are of
two forms: firstly, the matter fields contribute local quadratic
energy-momentum corrections via the tensor $S_{\mu\nu}$, and
secondly, there are nonlocal effects from the free gravitational
field in the bulk, transmitted via the projection of the bulk Weyl tensor, ${\cal
E}_{\mu\nu}$. The local matter corrections are
given by
\begin{equation}
S_{\mu\nu}=\frac{1}{12}T_\alpha{}^\alpha T_{\mu\nu}
-\frac{1}{4}T_{\mu\alpha}T^\alpha{}_\nu+
\frac{1}{24}g_{\mu\nu} \left[3 T_{\alpha\beta}
T^{\alpha\beta}-\left(T_\alpha{}^\alpha\right)^2 \right]\,,
\end{equation}
which is equivalent to
\begin{equation}
S_{\mu\nu}^{\ \ perfect\ fluid}=\frac{1}{12} \rho^2 u_\mu u_\nu
+\frac{1}{12}\rho\left(\rho+2 p\right)h_{\mu\nu}\,,
\end{equation}
for a perfect fluid 
and 
\begin{eqnarray}
S_{\mu\nu}^{\ \ scalar\ field}&=&\frac{1}{6}\left(-\frac{1}{2}\phi_{;\alpha}\phi^{;\alpha}
+V(\phi)\right)\phi_{;\mu}\phi_{;\nu}\nonumber\\
&&\qquad\qquad +\frac{1}{12}\left(-\frac{1}{2}\phi_{;\alpha}\phi^{;\alpha}
+V(\phi)\right)\left(-\frac{3}{2}\phi_{;\alpha}\phi^{;\alpha}
-V(\phi)\right)g_{\mu\nu}\,,
\end{eqnarray}
for a minimally-coupled scalar field.  If we have both a perfect fluid and a scalar field and if we  assume that the gradient of the scalar field $\phi^{;\,\mu}$, is aligned with the fluid 4-velocity, $u^{\mu}$, that is $\phi^{;\,\mu}/\sqrt{-\phi_{;\,\alpha}\phi^{;\,\alpha}}=u^\mu$. (In general $\phi^{;\,\mu}$ need not be aligned with $u^\mu$ thereby creating a rich variety of cross terms.) The local brane effects due to a combination of a perfect fluid and a scalar field are  then
\begin{eqnarray}
S_{\mu\nu} &=&\frac{1}{12} \left(\rho-\frac{1}{2}\phi_{;\,\alpha}\phi^{;\,\alpha}+V(\phi)\right)^2 u_\mu u_\nu\nonumber\\
&&+\frac{1}{12}\left(\rho-\frac{1}{2}\phi_{;\,\alpha}\phi^{;\,\alpha}+V\right)
\left(\rho+2 p-\frac{3}{2}\phi_{;\,\alpha}\phi^{;\,\alpha}-V(\phi)\right)h_{\mu\nu}
\end{eqnarray}

The non-local effects from the free gravitational field in the bulk are characterized by the projection of the bulk Weyl tensor onto the brane.  Given a timelike congruence on the brane, the bulk correction, ${\cal
E}_{\mu\nu}$ can be decomposed \cite{Maartens:2000a} via
\begin{equation}
{\cal E}_{\mu\nu}=-\left(\frac{\widetilde \kappa}{\kappa}\right)^4\left[{\cal U}(u_{\mu}u_{\nu}+\frac{1}{3}h_{\mu\nu})+{\cal P}_{\mu\nu}+ 2{\cal Q}_{(\mu}u_{\nu)}\right]
\end{equation}
(See \cite{Maartens:2000a} for further details.)  In general, the conservation equations (the contracted Bianchi identities on the brane) do not determine all of the independent components of ${\cal E}_{\mu\nu}$ on the brane. In particular, there is no
evolution equation for ${\cal P}_{\mu\nu}$ and hence, in general, the projection of the
5-dimensional field equations onto the brane does not lead to a
closed system.  However, in the cosmological context studied here, we will assume
\begin{equation}\label{bulk_weyl}
\D_\mu{\cal U}={\cal Q}_{\mu}={\cal P}_{\mu\nu}=0\,,
\end{equation}
where $\D_\mu$ is the totally projected part of the brane
covariant derivative.
Since ${\cal P}_{\mu\nu}=0$,
in this case the evolution of ${\cal E}_{\mu\nu}$ is fully determined \cite{randall}.
In general ${\cal U}={\cal U}(t)\neq0$ (and can be negative) in the Friedmann
background~\cite{cline,BinDefLan:2000}.
For a spatially homogeneous and isotropic
model on the brane, equation (\ref{bulk_weyl}) follows, and 
similar conditions apply self-consistently in other Bianchi models \cite{Maartens:2000a}.
In the appendix we prove that the integrability conditions for ${\cal P}_{\mu\nu}=0,{\cal Q}_{\mu}=0$ implies spatial homogeneity.  Physically ${\cal P}_{\mu\nu}$ corresponds to gravitational waves and will not affect the dynamics close to the singularity \cite{Langlois:2001}.  From the analysis of the evolution equation for ${\cal Q}_{\mu}$ close to the initial singularity, it can be shown that a small ${\cal Q}_{\mu}$ does not affect the dynamical evolution of $\cal U$ (to lowest order) \cite{Maartens:2000a,Gordon:2001,Coley:2002a}.

All of the bulk corrections mentioned above may be consolidated into an effective total
energy density and pressure as follows. The modified Einstein equations take the standard
Einstein form with a redefined energy-momentum tensor:
\begin{equation}
G_{\mu\nu}=\kappa^2 T^{\rm total}_{\mu\nu}\,,
\end{equation}
where
\begin{equation}
T^{\rm tot}_{\mu\nu} \equiv-\frac{\Lambda}{\kappa^2}g_{\mu\nu}+ T_{\mu\nu}+\frac{\widetilde{\kappa}^{4}}{\kappa^2}S_{\mu\nu}- \frac{1}{\kappa^2}{\cal E}_{\mu\nu} 
\end{equation}
is the redefined equivalent perfect fluid energy-momentum tensor
with the total equivalent energy density due to both a perfect fluid and a scalar field
\begin{eqnarray} 
\rho^{\rm total} \label{total1} &=&\frac{\Lambda}{\kappa^2}+ \rho+\left(-\frac{1}{2}\phi_{;\,\alpha}\phi^{;\,\alpha}+V(\phi)\right)
 \nonumber\\&& \qquad+\frac{\widetilde{\kappa}^{4}}{\kappa^6}
\Biggl[\frac{\kappa^4}{12}\left(\rho-\frac{1}{2}\phi_{;\,\alpha}\phi^{;\,\alpha}
+V(\phi)\right)^2+{\cal U}\Biggr] \\
p^{\rm total} \label{total2} &=& -\frac{\Lambda}{\kappa^2}+p+\left(-\frac{1}{2}\phi_{;\,\alpha}\phi^{;\,\alpha}-V(\phi)\right)\nonumber\\
&&+\frac{\widetilde{\kappa}^{4}}{\kappa^6}
\Biggl[ \frac{\kappa^4}{12}\left(\rho-\frac{1}{2}\phi_{;\,\alpha}\phi^{;\,\alpha}+V(\phi)\right)
\left(\rho+2p-\frac{3}{2}\phi_{;\,a}\phi^{;\,a}-V(\phi)\right)
+\frac{1}{3}{\cal U}\Biggr] 
\end{eqnarray}
where we have assumed that
$D_\mu{\cal U}={\cal Q}_{\mu}={\cal P}_{\mu\nu}=0$
in the cosmological case of interest here.

As a consequence of the form of the bulk energy-momentum tensor
and of $Z_2$ symmetry, it follows \cite{Shiromizu:2000} that
the brane energy-momentum tensor separately satisfies the
conservation equations, (where we have tacitly assumed that the scalar field and the matter are non-interacting) i.e.,
\begin{eqnarray}
T^{\mu\ \ perfect\ fluid}_{\nu;\,\mu}&=&0 {\rm\ \  which\  yields\ \ } \dot\rho+3H(\rho+p)=0\label{matter_conservation}\\
T^{\mu\ \ scalar\ field}_{\nu;\,\mu}&=&0 {\rm\ \  which\  yields\ \ } \ddot\phi+3H\dot\phi+\frac{\partial V}{\partial \phi}=0\label{Klein-Gordon}\,,
\end{eqnarray}
whence  the Bianchi identities on the brane imply that the projected
Weyl tensor obeys the constraint
\begin{equation}
{\cal E}^{\mu}_{\nu;\,\mu}=\widetilde{\kappa}^4 S^{\mu}_{\nu;\,\mu}{\rm \ \ which\  yields\ \ }
\dot{\cal U}+4H{\cal U}=0\label{Udot}
\end{equation}


\section{Bianchi IX Models}

\subsection{Setting up the Dynamical System}

We shall  use the formalism of Hewitt, Uggla, and Wainwright  introduced in \cite{Elliswainwright} for positive curvature models (see pages 179--182 in \cite{Elliswainwright}).  The source term (restricted to the Brane) is a non-interacting mixture of ordinary matter having energy density $\rho$, and a minimally coupled scalar field, $\phi$.  We shall assume that  the matter content is equivalent to that of a non-tilting perfect fluid with a linear barotropic equation of state for the fluid, i.e., 
$p = (\gamma-1)\rho$, where the energy conditions  impose the
restriction $\rho\geq 0$, and the constant $\gamma$ satisfies $\gamma\in[0,2]$
from causality requirements. We shall also assume that the scalar field potential has an exponential form $V(\phi)=V_0e^{ k\kappa\phi}=V_0e^{\frac{\sqrt{8\pi}}{M_p}k\phi}$ \cite{La89a,Acetta90,Salam84,Halliwell87}. 
The energy-momentum tensor describing the source is given in equation (\ref{source})
where, for a homogeneous scalar field, $\phi=\phi(t)$.   

Our variables are the same as those introduced by Hewitt, Uggla and Wainwright (see page 180 in \cite{Elliswainwright}), with the addition of
\begin{eqnarray}\label{new_vars}
\tilde\Omega &=&\frac{\kappa^2 \rho}{3D^2},\qquad\qquad 
\tilde\Omega_{\Lambda} = \frac{\Lambda}{3D^2},\qquad\qquad
\tilde\Phi=\frac{\kappa^2 V}{3D^2},\nonumber\\
&&\tilde\Psi=\sqrt{\frac{3}{2}}\frac{\kappa\dot\phi}{3D},\qquad\qquad
\tilde \Omega_{\cal U}= \frac{\cal U}{3D^2}
\end{eqnarray}
where $$ D\equiv \sqrt{H^2+\frac{1}{4}(n_1n_2n_3)^{2/3}}.$$
The total equivalent dimensionless energy density due to all sources and bulk corrections is
\begin{equation}
\tilde \Omega^{total}\equiv
\frac{\kappa^2\rho^{total}}{3D^2}
=\tilde\Omega
+\tilde\Omega_{\Lambda}
+\tilde\Phi+\tilde\Psi^2
+a^2\tilde\Omega_{\cal U}+\frac{a^2}{4}D^2\left( \tilde\Omega + \tilde\Phi+\tilde\Psi^2\right)^2,
\end{equation}
where $$a^2\equiv\frac{\widetilde \kappa^4}{\kappa^4}$$
 
The governing differential equations for the quantities $${\bf X}\equiv [D,\tilde H,\tilde\Sigma_1,\tilde\Sigma_2,\tilde  N_1,\tilde N_2,\tilde N_3, \tilde\Omega,\tilde\Omega_{\Lambda},\tilde\Psi,\tilde\Phi,\tilde\Omega_{\cal U}] $$
are as follows
\begin{subequations} \label{DSgeneral}
\begin{eqnarray}
         D'        &=&-(1+\tilde q)\tilde H D \label{D_prime}\\
  \tilde H '       &=& \tilde q (\tilde H^2-1) \label{H_prime}\\
  \tilde \Sigma_+' &=& (\tilde q-2)\tilde H\tilde\Sigma_+-\tilde S_+ \label{Sigma+_prime}\\
  \tilde \Sigma_-' &=& (\tilde q-2)\tilde H\tilde\Sigma_--\tilde S_- \label{Sigma-_prime}\\
  \tilde N_1 '     &=&  \tilde N_1(\tilde H\tilde q -4\tilde\Sigma_+) \label{N1_prime}\\
  \tilde N_2 '     &=&  \tilde N_2(\tilde H\tilde q 
                        +2\tilde\Sigma_++2\sqrt{3}\tilde\Sigma_-) \label{N2_prime}\\
  \tilde N_3 '     &=&  \tilde N_3(\tilde H\tilde q 
                        +2\tilde\Sigma_+-2\sqrt{3}\tilde\Sigma_-) \label{N3_prime}\\
  \tilde\Omega'    &=& \tilde H\tilde\Omega\left(2(\tilde q +1)-3\gamma\right) 
                    \label{Omega_prime} \\
  \tilde\Omega_{\Lambda}'    &=& 2\tilde H\tilde\Omega_{\Lambda}\left(\tilde q +1\right) \label{OmegaL_prime}\\
  \tilde\Psi'      &=&(\tilde q-2)\tilde H\tilde\Psi-\frac{\sqrt{6}}{2}k\tilde\Phi
                    \label{Psi_prime}\\
  \tilde\Phi'      &=&2\tilde\Phi\left((1+\tilde q)\tilde H
                         +\frac{\sqrt{6}}{2}k\tilde\Psi\right) \label{Phi_prime}  \\   
  \tilde\Omega_{\cal U} ' &=& 2\tilde H\tilde\Omega_{\cal U}(\tilde q -1) 
                    \label{Omega(U)_prime}  
\end{eqnarray}
\end{subequations} 
 
The quantity $\tilde q$ is the deceleration parameter, and $\tilde S_+$ and $\tilde S_-$ are curvature terms that are defined by the following expressions:  
\begin{eqnarray}
\tilde q &\equiv&   
2\tilde \Sigma_+^2
+2\tilde\Sigma_-^2
+ \frac{(3\gamma-2)}{2}\tilde\Omega
-\tilde\Omega_{\Lambda}
+2\tilde\Psi^2
-\tilde\Phi  \nonumber\\
&&+a^2\tilde\Omega_{\cal U}
 +\frac{a^2}{4}D^2\Biggl[\left( \tilde\Omega + \tilde\Psi^2+\tilde\Phi\right)
                        \left((3\gamma-1)\tilde\Omega + 5\tilde\Psi^2-\tilde\Phi\right)
                 \Biggr] \label{qdef}\\
{\tilde{S}_{+}}&\equiv&{\displaystyle \frac {1}{6}}\, \left( \! \,
        {\tilde{N}_{2}} - {\tilde{N}_{3}}\, \!  \right) ^{2} 
        - {\displaystyle \frac {1}{6}}\,{\tilde{N}_{1}}\, 
        \left( \! \,2\,{\tilde {N}_{1}} - {\tilde{N}_{2}} - 
        {\tilde{N}_{3}}\, \!  \right) \label{S+def}\\
{\tilde{S}_{-}}&\equiv& \frac{1}{6}\,\sqrt {3}\, \left( \! \,{\tilde{N}_{2}} -
         {\tilde{N}_{3}}\, \!  \right) \, \left( \! \,
         - {\tilde{N}_{1}} + {\tilde{N}_{2}} + {\tilde{N}_{3}}\, \!  \right) \label{S-def}
\end{eqnarray} 
 
The evolution equations for (\ref{Omega_prime}) comes from the conservation equation (\ref{matter_conservation}).  The evolution equations for (\ref{Psi_prime}) and (\ref{Phi_prime}) are derived from the Klein-Gordon equation derived from the conservation equation (\ref{Klein-Gordon}).  The evolution equation (\ref{Omega(U)_prime}) comes from the conservation equation (\ref{Udot}).

In addition there are two constraint equations that must also be satisfied,
\begin{subequations}
\begin{eqnarray}
G_1({\bf X})&\equiv& \tilde H^2 +\frac{1}{4}(\tilde N_1\tilde N_2 \tilde N_3)^{2/3} -1=0
                           \label{constraint1}\\
G_2({\bf X})&\equiv&1-  \tilde \Sigma_+^2-\tilde \Sigma_-^2-\tilde\Omega^{total} -\tilde V=0\label{constraint2}
\end{eqnarray}
\end{subequations}
where$$\tilde V =\frac{1}{12}\left[({\tilde N_1}^2+{\tilde N_2}^2+{\tilde N_3}^2-2(\tilde N_1\tilde N_2 + \tilde N_1\tilde N_3+\tilde N_2\tilde N_3)+3(\tilde N_1\tilde N_2\tilde N_3)^{2/3}\right].$$ 
Equation (\ref{constraint1}) follows from the definition of $D$, and equation (\ref{constraint2}) is the generalized Friedmann equation.
We now have determined the equations describing the evolution of the Bianchi type IX brane world models.  The resulting equations are suitable for a qualitative analysis using techniques from dynamical systems theory. In general, the system of equations (\ref{DSgeneral}) can be interpreted as ${\bf X}' = {\bf F}({\bf X})$ where ${\bf F}:{\bf X}\in {\mathbb R}^{12} \to {\mathbb R}^{12}$.  We must also make careful note of the two constraint equations $G_{1}({\bf X})=0$ and $G_{2}({\bf X})=0$.  These constraint equations essentially restrict the dynamics of the dynamical system ${\bf X}' = {\bf F}({\bf X})$ to lower dimensional surfaces in ${\mathbb R}^{12}$.  In principal, these constraint equations may be used to eliminate one of the twelve variables provided the constraint is not singular.

\subsection{Symmetry Transformations}

Since the dynamical system (\ref{DSgeneral}) is invariant under the transformation $\tilde \Phi \to -\tilde\Phi$ we can restrict our state space to $D\geq 0$ [by definition of $D$, see Eq. (\ref{new_vars})] and $\tilde \Phi \geq 0$ without loss of generality. 

Note the dynamical system (\ref{DSgeneral}) is also invariant under the transformation 
\begin{equation}
(\tilde \Sigma_+,\tilde \Sigma_-,\tilde N_1, \tilde N_2, \tilde N_3)\to (-\frac{1}{2}\Sigma_+-\frac{\sqrt{3}}{2}\Sigma_-, \frac{\sqrt{3}}{2}\tilde \Sigma_+-\frac{1}{2}\tilde \Sigma_-,\tilde N_2, \tilde N_3,
\tilde N_1).\label{transformation}
\end{equation}
This symmetry implies that any equilibrium point with a non-zero $\tilde \Sigma_{\pm}$ term, will have two equivalent copies of that point located at positions that are rotated thru an angle of $2\pi/3$ and centered along a different axis of the $N_{\alpha}$.

\subsection{Invariant Sets}

If we assume the weak energy condition for a perfect fluid (i.e., $\rho\geq 0$), then we must restrict the state space to the set of points $\tilde \Omega\geq 0$.  Since we are investigating the behaviour of the Bianchi Type IX brane world models in particular, we can restrict the state space to $N_\alpha\geq 0$ without loss of generality.
Therefore the state space for the Bianchi IX brane world models is the set of points ${\cal S}=\{ {\bf X}\in {\mathbb R}^{12}| G_{1}({\bf X})=0, G_{2}({\bf X})=0,\tilde \Omega\geq 0,{\tilde N}_\alpha\geq 0, D\geq 0,\tilde\Phi \geq 0\}$ .

The evolution equation for $\tilde\Omega_{\cal U}$, (\ref{Omega(U)_prime}) implies that the surface $\tilde\Omega_{\cal U}=0$ divides the state space into three distinct regions,
${\cal U}^{+}=\{{\bf X}\in {\cal S}|\tilde\Omega_{\cal U} > 0\}$, ${\cal U}^{0}=\{{\bf X}\in {\cal S}|\tilde\Omega_{\cal U} = 0\}$, and ${\cal U}^{-}=\{{\bf X}\in {\cal S}|\tilde\Omega_{\cal U} < 0\}$. 

There are various invariant sets associated with the matter content.  We define six {\em Matter invariant sets} as
\begin{eqnarray*}
^0\Omega^{0,0}&=&\{{\bf X}\in {\cal S}| \tilde\Omega=0, \tilde\Phi=0, \tilde\Psi=0 \},\\
^0\Omega^{0,\pm}&=&\{{\bf X}\in {\cal S}|\tilde\Omega=0, \tilde\Phi=0, \tilde\Psi\not =0 \},\\
^0\Omega^{+,\pm}&=&\{{\bf X}\in {\cal S}|\tilde\Omega=0, \tilde\Phi\not =0, \tilde\Psi\not = 0 \},\\
^+\Omega^{0,0}&=&\{{\bf X}\in {\cal S}|\tilde\Omega\not=0, \tilde\Phi=0, \tilde\Psi=0 \},\\
^+\Omega^{0,\pm}&=&\{{\bf X}\in {\cal S}|\tilde\Omega\not=0, \tilde\Phi=0, \tilde\Psi\not =0 \},\\
^+\Omega^{+,\pm}&=&\{{\bf X}\in {\cal S}|\tilde\Omega\not=0, \tilde\Phi\not =0, \tilde\Psi\not = 0 \},
\end{eqnarray*}
where the notation is interpreted as  $$ ^{({\rm value\ of\ } \tilde \Omega)} \Omega ^{({\rm value\ of\ }\tilde\Phi,{\rm value\ of\ }\tilde\Psi)}$$

In addition to the matter invariant sets, there are invariant sets associated with the geometry of the spacetime.  The dynamical system (\ref{DSgeneral}) implies that any combination of the conditions $\tilde N_\alpha>0$ and $\tilde N_\beta=0$ defines an invariant set.  Since these conditions yield the Bianchi type of the underlying geometry we call these invariant sets, {\em Bianchi invariant sets}.  The Bianchi type corresponding to the different combinations are 
\begin{eqnarray*}
B(I) &=& \{ {\bf X}\in {\cal S} | \tilde N_1=0, \tilde N_2=0,\tilde N_3=0 \} \\
B(II) &=& \{ {\bf X}\in {\cal S} | \tilde N_1\not =0, \tilde N_2=0,\tilde N_3=0 \} \\
B(VII_0) &=& \{ {\bf X}\in {\cal S} | \tilde N_1 =0, \tilde N_2\not=0,\tilde N_3\not=0 \} \\
B(IX) &=& \{ {\bf X}\in {\cal S} | \tilde N_1\not =0, \tilde N_2\not=0,\tilde N_3\not=0 \} 
\end{eqnarray*}
where the sets $B(II)$ and $B(VII_0)$ have two additional and equivalent disjoint copies of themselves that are determined by the transformation (\ref{transformation}).

From (\ref{constraint1}) we have that $-1\leq \tilde H \leq 1$ and $\tilde N_1 \tilde N_2 \tilde N_3\leq 8$.   Furthermore, in the invariant sets ${\cal U}^+$ and ${\cal U}^0$, using equation (\ref{constraint2}) it can be shown that 
$$0\leq \tilde \Sigma_+^2,\tilde \Sigma_-^2,\tilde\Omega,\tilde\Omega_{\Lambda},\tilde\Psi^2, \tilde \Phi ,\tilde V\leq 1.$$  However, knowing that $0\leq \tilde V\leq 1$ and $0\leq \tilde N_1 \tilde N_2 \tilde N_3\leq 8$ is not sufficient to place any bounds on the $N_\alpha$'s or $D$.  Furthermore, in the invariant set ${\cal U}^-$, we cannot place upper bounds on any of the variables without some redefinition of the dimensionless variables (\ref{new_vars}).

\subsection{First Integrals}

It is possible to show that the function
\begin{equation}
W=(\tilde H^2-1)^{[-2s-3\gamma t]}\, \tilde\Omega_\Lambda^{[s+(3\gamma-2)t]}\,\tilde\Omega^{2t}\,\tilde\Omega_{\cal U}^s
\end{equation}
(where $s,t$ are parameters that can take any value)
is a first integral of the of the dynamical system (\ref{DSgeneral}), that is $W'=0$.
For the particular values $s$ and $t$ we obtain the following invariants (for any value $K$),
\begin{eqnarray}
(s=0,t=1),&\qquad& \qquad K(\tilde H^2-1)^{3\gamma}={\tilde\Omega_{\Lambda}}^{(3\gamma-2)} \tilde\Omega ^2\\
(s=1,t=0),&\qquad&\qquad K(\tilde H^2-1)^{2}={\tilde\Omega_{\Lambda}} \tilde\Omega_{\cal U}\\
(s=-3\gamma,t=2), & \qquad &\qquad K {\tilde\Omega_{\cal U}}^{3\gamma} ={\tilde\Omega_{\Lambda}}^{(3\gamma-4)}\tilde\Omega^4 \\
(s=-(3\gamma-2),t=1),&\qquad&\qquad 
K{\tilde\Omega_{\cal U}}^{(3\gamma-2)}=(\tilde H^2-1)^{(3\gamma-4)}\tilde \Omega^2
\end{eqnarray}
In the invariant set ${\cal U}^+ \cup {\cal U}^0$, it is possible to show that $q\geq-1$, which implies that $\Omega_\Lambda \rightarrow 0$ as $\tau \rightarrow \infty$.  Using the first and second invariants, we easily obtain the result that $\tilde H^2\rightarrow 1$.
From here we also obtain the result that if $\gamma>4/3$ that $\tilde \Omega_{\cal U}\rightarrow 0$.


\section{Qualitative Analysis of the case $\tilde\Omega=\tilde\Omega_\Lambda=0$ }

In the invariant set ${\cal U}^+ \cup {\cal U}^0$, we can show that $q\geq-1$.  This implies that the invariant set $\Omega_{\Lambda}=0$ is the invariant set containing all of the past asymptotic behaviour (i.e, early times) for all ever-expanding models in ${\cal U}^+ \cup {\cal U}^0$.  It can be argued that a scalar field becomes essentially massless as it evolves backwards in time, hence it will dominate the dynamics at early times (see Section \ref{General_Dynamics}).   In our effort to understand the dynamical behaviour at early times we shall assume in the analysis that follows that there is no perfect fluid (${\tilde \Omega=0}$) and the four-dimensional cosmological constant is zero (${\tilde\Omega_{\Lambda}=0}$).

If $\tilde\Omega=\tilde\Omega_\Lambda=0$ then the equations (\ref{DSgeneral}) reduce to 
\begin{subequations}\label{DS_Scalar_Field}
\begin{eqnarray}
         D'        &=&-(1+\tilde q)\tilde H D  \label{SF_D_prime}\\
  \tilde H '       &=& \tilde q (\tilde H^2-1)  \label{SF_H_prime}\\
  \tilde \Sigma_+' &=& (\tilde q-2)\tilde H\tilde\Sigma_+-\tilde S_+  \label{SF_Sigma+_prime}\\
  \tilde \Sigma_-' &=& (\tilde q-2)\tilde H\tilde\Sigma_--\tilde S_-  \label{SF_Sigma-_prime}\\
  \tilde N_1 '     &=&  \tilde N_1(\tilde H\tilde q -4\tilde\Sigma_+)  \label{SF_N1_prime}\\
  \tilde N_2 '     &=&  \tilde N_2(\tilde H\tilde q 
                        +2\tilde\Sigma_++2\sqrt{3}\tilde\Sigma_-)  \label{SF_N2_prime}\\
  \tilde N_3 '     &=&  \tilde N_3(\tilde H\tilde q 
                        +2\tilde\Sigma_+-2\sqrt{3}\tilde\Sigma_-)  \label{SF_N3_prime}\\
  \tilde\Psi'      &=&(\tilde q-2)\tilde H\tilde\Psi-\frac{\sqrt{6}}{2}k\tilde\Phi
                    \label{SF_Psi_prime}\\
  \tilde\Phi'      &=&2\tilde\Phi\left((1+\tilde q)\tilde H
                         +\frac{\sqrt{6}}{2}k\tilde\Psi\right) \label{SF_Phi_prime}  \\   
  \tilde\Omega_{\cal U} ' &=& 2\tilde H\tilde\Omega_{\cal U}(\tilde q -1) 
                    \label{SF_Omega(U)_prime}  
\end{eqnarray}
\end{subequations} 
 
The quantity $\tilde q$ is the deceleration parameter, and $\tilde S_+$ and $\tilde S_-$ are curvature terms that are now defined by the following expressions:   
\begin{eqnarray}
\tilde q &\equiv&   2\tilde \Sigma_+^2+2\tilde\Sigma_-^2+2\tilde\Psi^2-\tilde\Phi \nonumber\\
&&+{a^2}\Biggl[\frac{1}{4}D^2(\tilde\Psi^2+\tilde\Phi)(5\tilde\Psi^2-\tilde\Phi)
+\tilde\Omega_{\cal U}\Biggr]\\
{\tilde{S}_{+}}&\equiv&{\displaystyle \frac {1}{6}}\, \left( \! \,
        {\tilde{N}_{2}} - {\tilde{N}_{3}}\, \!  \right) ^{2} 
        - {\displaystyle \frac {1}{6}}\,{\tilde{N}_{1}}\, 
        \left( \! \,2\,{\tilde {N}_{1}} - {\tilde{N}_{2}} - 
        {\tilde{N}_{3}}\, \!  \right) \\
{\tilde{S}_{-}}&\equiv& \frac{1}{6}\,\sqrt {3}\, \left( \! \,{\tilde{N}_{2}} -
         {\tilde{N}_{3}}\, \!  \right) \, \left( \! \,
         - {\tilde{N}_{1}} + {\tilde{N}_{2}} + {\tilde{N}_{3}}\, \!  \right)
\end{eqnarray} 
The two constraint equations  are
\begin{subequations}
\begin{eqnarray}
     1    &=&   \tilde H^2 +\frac{1}{4}(\tilde N_1\tilde N_2 \tilde N_3)^{2/3} 
                           \label{SFconstraint1}\\
\tilde H^2&=&   \tilde \Sigma_+^2+\tilde \Sigma_-^2+ \tilde\Omega^{total} +\frac{1}{12}\left((\tilde N_1^2+\tilde N_2^2+\tilde N_3^2)
        -2(\tilde N_1\tilde N_2+\tilde N_2\tilde N_3+\tilde N_1\tilde N_3)
                   \right) \label{SFconstraint2}
\end{eqnarray}
\end{subequations}
where 
\begin{equation}
\tilde \Omega^{total}\equiv\frac{\kappa^2\rho^{total}}{3D^2}= \tilde\Phi+\tilde\Psi^2+ a^2\Biggl[\frac{1}{4}D^2(\tilde\Phi+\tilde\Psi^2)^2+\tilde\Omega_{\cal U}\Biggr].
\end{equation}

\subsection{Equilibrium Points at Finite $D$}

Here we define $\bar {\bf X} = [D,\tilde H,\tilde \Sigma_+,\tilde \Sigma_-,\tilde N_1,\tilde N_2,\tilde N_3,\tilde \Psi, \tilde\Phi, \tilde \Omega_{\cal U}]$ and we restrict the state space accordingly to be $\bar {\cal S} = \{ {\bf X}\in {\cal S} | \tilde\Omega=0, \tilde\Omega_{\Lambda}=0\}$.
The equilibrium points $\bar {\bf X}_0$, can be classified into one of the three matter invariant sets that do not have a perfect fluid component. Note, $\epsilon$ is a discrete parameter where $\epsilon =1$ corresponds to expanding models, while $\epsilon = -1$ corresponds to contracting models.  The notation $[\times m]$ signifies that the preceding eigenvalue has multiplicity $m$.

\subsubsection{Vacuum, $^0\Omega^{0,0}$}

\begin{description}
 
\item {$R_{\epsilon}$; ROBERTSON-WALKER (Radiation) }\hfill\hfill\linebreak    
$\bar {\bf X }_0
=[0,\epsilon,0,0,0,0,0,\frac{1}{a^2}]$
The eigenvalues in the nine dimensional phase space [$\tilde \Omega_{\cal U}$ eliminated via Eq. (\ref{SFconstraint2})] are
$$\epsilon(-2,-1,-1,-1,1,1,1,2,4).$$  This point is obviously a saddle with a five dimensional unstable manifold ($\epsilon=1$), and hence when the dynamics are restricted to the remaining constraint surface Eq. (\ref{SFconstraint1})], this point will remain a saddle (It can be shown that the stable manifold on the remaining constraint surface has dimension four) .

\item{${R^{II}}_{\epsilon}$; BIANCHI II (Radiation)}   \hfill\hfill\linebreak 
$\bar {\bf X}_0
=[0,\epsilon,\epsilon\frac{1}{4},0,\frac{\sqrt{3}}{2},0,0,0,0,\frac{7}{8a^2}]$ 
and two other equivalent points obtained through transformation (\ref{transformation}).
The eigenvalues in the nine dimensional phase space [$\tilde \Omega_{\cal U}$ eliminated via Eq. (\ref{SFconstraint2})] are
$$\epsilon(-2,-1,-1,-\frac{1}{2}(1\pm\sqrt{6}i),\frac{3}{2},\frac{3}{2},2,4)$$  This point is obviously a saddle with a four dimensional unstable manifold ($\epsilon=1$), and hence when the dynamics are restricted to the remaining constraint surface Eq. (\ref{SFconstraint1})], this point will remain a saddle.

\item{${K}_{\epsilon}$; KASNER (Vacuum)} \hfill\hfill\linebreak 
$\bar {\bf X}_0=
[0,\epsilon,\cos(\theta),\sin(\theta),0,0,0,0,0,0]$ where $-\pi<\theta\leq\pi$.  
This is a special case of the Kasner surface defined in next section.  

\item{${B^{VII}}_{\epsilon}$; BIANCHI VII$_0$ (Vacuum)}\hfill\hfill\linebreak  
$\bar {\bf X}_0=
[0,\epsilon,-\epsilon,0,0,s,s,0,0,0]$ 
and two other equivalent lines of equilibria obtained through transformation (\ref{transformation}) where $0< s<\infty$.  As $s\to 0$, these lines of equilibria approach a point on ${K}_{\epsilon}$.  The eigenvalues in the nine dimensional phase space [$\tilde \Omega_{\cal U}$ eliminated via Eq. (\ref{SFconstraint2})] are $$\epsilon(-3,0,0,2,4,6,6,\pm 2si)$$  One of the zero eigenvalues corresponds to the fact that this is a one-dimensional set of equilibrium points. The span of eigenvectors associated with the eigenvalues $[0,2,4,6,+2si,-2si]$ does not include the `$D$' direction.  The eigenvector associated with the eigenvalue $-3$ is the only eigenvector with a `$D$' component.  Therefore this point is a saddle ($\epsilon=1$) with a one dimensional stable manifold in the eight dimensional phase space $\bar {\cal S}$  when the dynamics are restricted to the remaining constraint surface Eq. (\ref{SFconstraint1}). 
\end{description}

\subsubsection{Massless Scalar Field, $^0\Omega^{0,\pm}$}

\begin{description}
\item {${\cal K}_{\epsilon}$; KASNER (Massless Scalar Field)} \hfill\hfill\linebreak 
$\bar {\bf X}_0
 =[0,\epsilon,\sin(\varphi)\cos(\theta),\sin(\varphi)\sin(\theta),0,0,0,\cos(\phi),0,0]$ where $-\pi<\theta\leq\pi$ and $-\frac{\pi}{2}\leq \varphi \leq \frac{\pi}{2}$. 
Or $\tilde\Sigma_+^2+\tilde\Sigma_-^2+\tilde\Psi^2=1$.
The eigenvalues in the nine dimensional phase space [$\tilde \Omega_{\cal U}$ eliminated via Eq. (\ref{SFconstraint2})] are
$$\epsilon(-3,0,0,4,4){\rm \ and\ }2\epsilon-4\tilde\Sigma_+,2\epsilon+2\tilde \Sigma_+\pm2\sqrt{3}\tilde\Sigma_-,6\epsilon+\sqrt{6}k\tilde\Psi$$
The two zero eigenvalues correspond to the fact that this is a two-dimensional set of equilibrium points. It can be shown (as in the ${B^{VII}}_{\epsilon}$ case) that the eigenvalue $-3$ corresponds to the $D$ direction, therefore, this set of equilibria will always be a a set of saddle points with an unstable manifold having dimension no more than seven.
\end{description}

\subsubsection{Massive Scalar Field, $^0\Omega^{+,\pm}$}

\begin{description}
\item{${{\cal F}}^{+}_{\epsilon}$; ROBERTSON-WALKER (Positive Curvature, Scalar Field) } \hfill\hfill\linebreak 
$\bar {\bf X}_0
=[0,\frac{k\epsilon\sqrt{2}}{2},0,0,\sqrt{4-2k^2},\sqrt{4-2k^2},\sqrt{4-2k^2},-\frac{\epsilon\sqrt{3}}{3},\frac{2}{3},0]$
The point only exists for $k^2<2$ and as $k^2\to 2$ this point approaches $P_{\epsilon}$.   This is the only equilibrium point for finite values of $D$ for which we can directly calculate the eigenvalues of the dynamical system restricted to the eight dimensional phase space $\bar {\cal S}$. The eigenvalues in the eight dimensional phase space [$H,\tilde \Omega_{\cal U}$ eliminated via Eqs. (\ref{SFconstraint1} and \ref{SFconstraint2})] are
$$\epsilon(-\frac{\sqrt{2}}{2}k,0,-\frac{\sqrt{2}}{2}(k\pm\sqrt{k^2+8(k^2-2)}) [\times 2],-\frac{\sqrt{2}}{2}(k\pm\sqrt{k^2+4(2-k^2)}))$$  This point when it exists ($\epsilon=1$) has a six dimensional stable manifold, a one dimensional unstable manifold, and a one dimensional center manifold.    

\item{$P_{\epsilon}$; ROBERTSON-WALKER (Zero Curvature, Power Law Inflation) }\hfill\hfill\linebreak 
$\bar {\bf X}_0=[0,\epsilon,0,0,0,0,0,-\frac{k\epsilon\sqrt{6}}{6},1-\frac{k^2}{6},0]$
The eigenvalues in the nine dimensional phase space [$\tilde \Omega_{\cal U}$ eliminated via Eq. (\ref{SFconstraint2})] are
$$\epsilon(-\frac{k^2}{2},k^2-2,\frac{1}{2}(k^2-2) [\times 3],\frac{1}{2}(k^2-6)[\times 3],k^2-4)$$  
If $k^2<2$ ($\epsilon=1$), then this point is a local sink in the eight dimensional phase space $\bar {\cal S}$.
If $k^2<2$ ($\epsilon=-1$), then this point is a local source in the eight dimensional phase space $\bar {\cal S}$.  When $2<k^2<4$ and when $4<k^2<6$ this point is a saddle in the nine dimensional phase space ($\epsilon=1$), and hence when the dynamics are restricted to the remaining constraint surface Eq. (\ref{SFconstraint1}), this point will remain a saddle.  When $k^2=4$ this point experiences a bifurcation with the point ${\cal RSF}^{0}_\epsilon$, the Robertson-Walker radiation-scalar field scaling models.  When $k^2=6$ this point becomes part of the Kasner massless scalar field models ${\cal K}_\epsilon$.

\item{${RSF}^{0}_{\epsilon}$; ROBERTSON-WALKER  (Zero Curvature - Radiation-Scalar Field Scaling Model)}\hfill\hfill\linebreak 
$\bar {\bf X}_0=[0,\epsilon,0,0,0,0,0,-\frac{2\epsilon\sqrt{6}}{3k},\frac{4}{3k^2},\frac{k^2-4}{k^2a^2}]$
When $k^2>4$ this point is an element of ${\cal U}^+$ and when $k^2<4$ this point is an element of ${\cal U}^-$.  The eigenvalues in the nine dimensional phase space [$\tilde \Omega_{\cal U}$ eliminated via Eq. (\ref{SFconstraint2})] are
$$\epsilon(-2,-1,-1,1,1,1,2,-\frac{1}{2k}(k\pm\sqrt{k^2+16(4-k^2)}))$$  This point ($\epsilon=1$) is easily seen to be a saddle with a four dimensional unstable manifold when $k^2>4$,  and a five dimensional unstable manifold when $k^2<4$, and hence when the dynamics are restricted to the remaining constraint surface Eq. (\ref{SFconstraint1}), this point will remain a saddle in all cases.  Note, the cosmological model represented by this equilibrium point has the property that the energy density attributed to the scalar field is proportional to the energy density of the dark radiation coming from the bulk 
$$ \rho_{sf}=\frac{1}{2}(\dot\phi)^2+V(\phi) \propto {\cal U}=\rho_{\cal U}.$$

\item{${RSF^{II}}_{\epsilon}$; BIANCHI II ($\tilde\Omega_{\cal U}\not =0$), Radiation-Scalar Field Scaling Model }\hfill\hfill\linebreak 
$\bar {\bf X}_0=[0,\epsilon,\frac{\epsilon}{4},0,\frac{\sqrt{3}}{2},0,0,-\frac{2\epsilon\sqrt{6}}{3k},\frac{4}{3k^2},\frac{7k^2-32}{8k^2a^2}]$
and two other equivalent points obtained through transformation (\ref{transformation}).
The eigenvalues in the nine dimensional phase space [$\tilde \Omega_{\cal U}$ eliminated via Eq. (\ref{SFconstraint2})] are
\begin{eqnarray*}
&&\epsilon\Biggl(-2,-1,\frac{3}{2},\frac{3}{2},2,\\
&&\frac{1}{4k^2}\left[-2k^2\pm\sqrt{(2k^2)^2
-2k^2\left[(23k^2-64)\pm\sqrt{(23k^2-64)^2-64k^2(7k^2-32)}\right]}\right]\Biggr)\end{eqnarray*}
This point is always a saddle point in the nine dimensional phase space, and hence when the dynamics are restricted to the remaining constraint surface Eq. (\ref{SFconstraint1}), this point will remain a saddle in all cases.  Note, the cosmological model represented by this equilibrium point has the property that the energy density attributed to the scalar field is proportional to the energy density of the dark radiation coming from the bulk 
$$ \rho_{sf}=\frac{1}{2}(\dot\phi)^2+V(\phi) \propto {\cal U}=\rho_{\cal U}.$$

\item{${SF^{II}}_{\epsilon}$; BIANCHI II ($\tilde\Omega_{\cal U}=0$), Scalar Field Model }\hfill\hfill\linebreak
$\bar {\bf X}_0=[0,\epsilon,\frac{-2\epsilon(2-k^2)}{k^2+16},0,6\sqrt{\frac{(2-k^2)(k^2-8)}{(k^2+16)^2}},0,0,-\frac{\epsilon k 3\sqrt{6}}{k^2+16},36\frac{8-k^2}{(k^2+16)^2},0]$ and two other equivalent points obtained through transformation (\ref{transformation}).  This point only exists when $2<k^2<8$.  As $k^2\to 2$ this point approaches the point $P_\epsilon$ and as $k^2\to 8$ this point approaches a point in the set ${\cal K}_\epsilon$.
The eigenvalues in the nine dimensional phase space [$\tilde \Omega_{\cal U}$ eliminated via Eq. (\ref{SFconstraint2})] are
\begin{eqnarray*}
&&\frac{\epsilon}{k^2+16}\Biggl(6(k^2-8)[\times 2],16(k^2-2),12(k^2-2)[\times 2],2(7k^2-32),-9k^2,\\
&&3[(k^2-8)\pm\sqrt{(k^2-8)^2+12(k^2-2)(k^2-8)}]\Biggr)
\end{eqnarray*}
This point is always a saddle point in the nine dimensional phase space, and hence when the dynamics are restricted to the remaining constraint surface Eq. (\ref{SFconstraint1}), this point will remain a saddle in all cases. The cosmological model represented by this equilibrium point is that of a Bianchi II model with a exponential potential scalar field.
\end{description}

Note there are other equilibrium points to the dynamical system (\ref{DS_Scalar_Field}), but they correspond to points that are neither inside the set $\bar {\cal S}$ nor on its boundary, and do not represent any limiting behaviour of the Bianchi Type IX brane-world scalar-field models.

\subsection{Initial singularity}

In the analysis of local equilibrium points at finite $D$ we found in the previous subsection that there were no local sources corresponding to expanding models. In order to complete the analysis, and determine the dynamical behaviour close to the initial singularity, we need to examine what happens as $D\rightarrow \infty$.  In this subsection we present a heuristic analysis of the situation, and include ordinary matter, that is $\tilde\Omega \not = 0$.

We define a new bounded variable
\begin{equation}
d=\frac{D}{D+1}, \ \ \ \ \ \ \ \ \ \ -1\leq d\leq 1
\end{equation}
and examine what happens as $d \rightarrow 1$ (assuming $\tilde{H}>0$). From Eqs. (\ref{D_prime} and \ref{H_prime}) (or 
(\ref{SF_D_prime})
 and \ref{SF_H_prime})) 
we have that $d \rightarrow 1$ 
and $ \tilde{H}\rightarrow 1$ 
monotonically (see also Section \ref{General_Dynamics}) and hence we need to consider the equilibrium points in the set $d=1$.

The analysis depends on whether the quantity 
\begin{equation}
{\cal A}= \frac{a^2}{4}(\tilde{\Omega}+\tilde{\Psi}^2+\tilde{\Phi})\left((3\gamma-1)\tilde{\Omega}+5\tilde{\Psi}^2-\tilde{\Phi}\right)
\end{equation}
in the expression for $\tilde q$ in equation (\ref{qdef}) is zero or not in an open neighborhood of the singularity. If ${\cal A}\not=0$, 
and assuming ${\cal A}>0$   (see Section \ref{General_Dynamics}), we define a new time variable by
\[ \dot{f}=\frac{(1-d)^2}{{\cal A}\tilde{H}}f'\]
and the remaining evolution eqns (on $d=1$) become
\begin{eqnarray}
\dot{\tilde{\Sigma}}_+&=&\tilde{\Sigma}_+,\quad
\dot{\tilde{\Sigma}}_-=\tilde{\Sigma}_-,\quad
\dot{\tilde{N}}_1=\tilde{N}_1,\quad
\dot{\tilde{N}}_2=\tilde{N}_2,\quad
\dot{\tilde{N}}_3=\tilde{N}_3,\nonumber\\
\dot{\tilde{\Omega}}&=&2\tilde{\Omega},\quad
\dot{\tilde{\Omega}}_\Lambda=2\tilde{\Omega}_\Lambda,\quad
\dot{\tilde{\Psi}}=\tilde{\Psi},\quad
\dot{\tilde{\Phi}}=2\tilde{\Phi},\quad
\dot{\tilde{\Omega}}_{\cal U}=2\tilde{\Omega}_{\cal U}.
\end{eqnarray}
Therefore, the only equilibrium point is 
\[\tilde{\Sigma}_+=\tilde{\Sigma}_-=\tilde{N}_1=\tilde{N}_2=\tilde{N}_3=\tilde{\Omega}=\tilde{\Omega}_\Lambda=\tilde{\Psi}=\tilde{\Phi}=\tilde{\Omega}_{\cal U}=0\]
and this is a local source. This equilibrium point corresponds to the Brane-Robertson-Walker solution \cite{Coley:2002a} with 
\[\frac{a^2}{4}D^2\tilde{\Psi}^2=1, \qquad\qquad D^2\tilde{\Phi}=0 \qquad (D^2\tilde{\Omega}=0)\]
(see Section \ref{General_Dynamics}).

It remains to consider the case ${\cal A}=0$. However in this case 
\begin{equation}
\tilde{q}=2(\tilde{\Sigma}_+^2+\tilde{\Sigma}_-^2)+2\tilde{\Psi}^2-\tilde{\Phi}+\frac{3\gamma-2}{2}\tilde{\Omega}
-\tilde{\Omega}_\Lambda+a^2\tilde{\Omega}_{\cal U}
\end{equation}
and resulting dynamical system (\ref{SF_H_prime}) - (\ref{SF_Omega(U)_prime}) has the same local equilibrium points as in the previous subsection but with $d=1$ (instead of $d=0$). In particular, there are no local sources, and the Kasner equilibrium points are saddles \cite{Coley:2002a}. Hence all orbits asymptote in the past to the BRW source. This will be discussed in more detail in the next section.

\subsection{Local stability of the Brane Robertson-Walker solution}

In order to analyze the dynamical system (\ref{DS_Scalar_Field}) for large values of $D$, we can define the following new variables
$$\tilde \Phi = r^2\sin^2\theta,\qquad {\rm and}\qquad \tilde\Psi=r\cos\theta,\qquad \tilde\Omega_\lambda=\frac{1}{4}a^2D^2(\tilde\Phi+\tilde\Psi^2)^2=\frac{1}{4}a^2D^2r^4$$
Using these new cylindrical coordinates, the infinite variable $D$ is essentially replaced by the bounded variable $\tilde \Omega_{\lambda}$ in the set ${\cal U}\cup{\cal U}^0$, hence in this set the only variables that remain unbounded are the $\tilde N_\alpha$'s.

The dynamical system (\ref{DS_Scalar_Field}) becomes 
\begin{subequations}\label{DS_Scalar_Field_Infinity}
\begin{eqnarray}
  \tilde H '       &=& \tilde q (\tilde H^2-1) \\
  \tilde \Sigma_+' &=& (\tilde q-2)\tilde H\tilde\Sigma_+-\tilde S_+\\
  \tilde \Sigma_-' &=& (\tilde q-2)\tilde H\tilde\Sigma_--\tilde S_-\\
  \tilde N_1 '     &=&  \tilde N_1(\tilde H\tilde q -4\tilde\Sigma_+)\\
  \tilde N_2 '     &=&  \tilde N_2(\tilde H\tilde q 
                        +2\tilde\Sigma_++2\sqrt{3}\tilde\Sigma_-)\\
  \tilde N_3 '     &=&  \tilde N_3(\tilde H\tilde q 
                        +2\tilde\Sigma_+-2\sqrt{3}\tilde\Sigma_-)\\
  \tilde\Omega_{\cal U} ' &=& 2\tilde H\tilde\Omega_{\cal U}(\tilde q -1) 
                    \\
  \tilde\Omega_{\lambda}' &=& 2\tilde H\tilde\Omega_{\lambda}(\tilde q +1 -6\cos^2\theta) \\                  
  r'    &=& r\tilde H (\tilde q-2+3\sin^2\theta)\\
  \theta' &=& \sin\theta(3\tilde H\cos\theta+\frac{\sqrt{6}}{2}k r)
\end{eqnarray}
\end{subequations} 
where $\tilde S_+$ and $\tilde S_-$ are curvature terms defined previously and where 
\begin{equation}
\tilde q \equiv   2\tilde \Sigma_+^2+2\tilde\Sigma_-^2+r^2(3\cos^2\theta-1) +\tilde\Omega_{\lambda}(6\cos^2\theta-1) +a^2\tilde\Omega_{\cal U} 
\end{equation} 
The two constraint equations become  
\begin{subequations}
\begin{eqnarray}
     1    &=&   \tilde H^2 +\frac{1}{4}(\tilde N_1\tilde N_2 \tilde N_3)^{2/3} \\
\tilde H^2&=&   \tilde \Sigma_+^2+\tilde \Sigma_-^2+ \tilde\Omega_{\lambda}+a^2\tilde\Omega_{\cal U}+r^2 \nonumber \\
&&\qquad +\frac{1}{12}\left((\tilde N_1^2+\tilde N_2^2+\tilde N_3^2)
        -2(\tilde N_1\tilde N_2+\tilde N_2\tilde N_3+\tilde N_1\tilde N_3)
                   \right) 
\end{eqnarray}
\end{subequations}
From the constraint equations, we have $0\leq \tilde H^2,{\tilde \Sigma_+}^2,{\tilde\Sigma_-}^2,\tilde\Omega_{\cal U},\tilde\Omega_{\lambda},r\leq 1$, $0\leq \theta\leq \pi$, and $\tilde N_1\tilde N_2\tilde N_3 \leq 8$ (assuming $\tilde \Omega_{\cal U}\geq 0$).

The Brane-Robertson Walker solution  \cite{BinDefLan:2000} is represented by an equilibrium point in the set $\tilde \Omega_{\lambda}\not = 0$ ($D\to \infty$).  If we let $\tilde {\bf X}=[\tilde H,\tilde\Sigma_+,\tilde\Sigma_-,\tilde N_1,\tilde N_2,\tilde N_3,\tilde\Omega_{\cal U},\tilde \Omega_{\lambda},r,\theta]$ then this new equilibrium point is 
\begin{description}
\item{$m_\epsilon$ ; Brane Robertson Walker}\hfill\hfill\linebreak 
$\tilde{\bf X}_0=[\epsilon,0,0,0,0,0,0,1,0,\frac{\pi}{2}\pm\frac{\pi}{2}]$
and 
$ \tilde{\bf X}_0=[\epsilon,0,0,0,0,0,0,1,0,\frac{\pi}{2}]$.   
Using the constraint equation to eliminate $\tilde\Omega_{\cal U}$, the eigenvalues of the linearization at the point $\theta=0,\pi$ are $$\epsilon(10,10,5,5,5,3,3,3,3)$$
and when $\theta = \frac{\pi}{2}$ the eigenvalues of the linearization are
$$\epsilon(-3,-3,-3,-2,-2,-1,-1,-1,0)$$

\end{description}

The values $\theta=0,\pi$ correspond to invariant directions in the neighborhood of this equilibrium point.  For example $\theta=0,\pi$ corresponds to $\tilde \Phi =0$ invariant surface (that is the massless scalar field models).  The value $\theta=\pi/2$ corresponds to $\tilde \Psi=0$ is not an invariant direction, and we note that $\theta' >0$  in a neighborhood of the equilibrium point near $\tilde \Psi=0$. We easily observe that this equilibrium point is a source that strongly repels away from $\tilde \Psi=0$.  That is, if one was to approach this equilibrium point along a typical orbit (backwards through time), one would asymptotically approach a massless scalar field Brane Robertson-Walker solution.

\begin{table}[h]
\caption{Local sinks and sources.  Note that each of these local sinks has a corresponding local source, $P_-$ and $m_+$. }\label{sink}
\begin{tabular}{|c|c|l|}
\hline
Sink  & Conditions on $k$  & Description  \\
\hline
$P_+$(Zero curvature, power law inflation) & $k<\sqrt{2}$ & $D=0$, $\tilde{H}=1$, expanding models. \\
\hline
$m_-$ (Zero curvature, isotropic Braneworld) & all $k$ & $D\to \infty$, $\tilde H = -1$, contracting models\\
\hline 
 \end{tabular}
\end{table}


\section{General Dynamics}\label{General_Dynamics}

We now return to the general case (i.e., equations (\ref{DSgeneral}) - (\ref{S-def})) and include a general potential for the scalar field, and study what happens as $D\rightarrow\infty$ at the initial singularity (assuming this occurs). We shall also assume normal matter with $1\leq\gamma < 2$. 

We recall that there are no sources for finite values of $D$. Also, for a general scalar field, equations (\ref{Psi_prime}) and (\ref{Phi_prime}) become 
\begin{eqnarray}
\tilde \Psi ' &=& (\tilde q-2)\tilde H\tilde \Psi - \bar \epsilon \tilde\Phi \\
\tilde \Phi ' &=& 2(\tilde q +1)\tilde H\tilde \Phi + 2 \bar\epsilon \tilde\Phi\tilde\Psi
\end{eqnarray}
where $\bar \epsilon$ (related to the usual inflationary slow roll parameter $\epsilon$) is defined $$\bar \epsilon \equiv \sqrt{\frac{3}{2}}\frac{V_{\phi}}{\kappa V}.$$

From the Friedmann equation we have that 
\begin{equation}
\frac{a^2}{4}D^2(\tilde{\Omega}+\tilde{\Phi}+\tilde{\Psi}^2)^2=\tilde{\Omega}_{\lambda}\leq 1
\end{equation}
and hence each term $D\tilde{\Omega}$, $D\tilde{\Phi}$, $D\tilde{\Psi}^2$ is bounded (since the left-hand side is the sum positive definite terms). Hence, as $D\rightarrow\infty$, $\tilde{\Omega}$, $\tilde{\Psi}^2$, $\tilde{\Phi}\rightarrow 0$.

It is easy to show that $\tilde{\Omega}_\Lambda$, $\tilde{\Omega}_u \rightarrow 0$ as $D\rightarrow \infty$. Hence (\ref{qdef}) becomes
\begin{equation}
\tilde{q}=2(\tilde{\Sigma}_+^2+\tilde{\Sigma}_-^2)+\frac{a^2}{4}D^2\left([\tilde{\Omega}+\tilde{\Psi}^2+\tilde{\Phi}][(3\gamma-1)\tilde{\Omega}+5\tilde{\Psi}^2-\tilde{\Phi}]\right)
\end{equation}
Assuming $\tilde{H}>0$, equations (\ref{D_prime}) and (\ref{H_prime}) imply that as $\tau\rightarrow -\infty$ for $D\rightarrow \infty$, either $\tilde{q}\rightarrow 0$ or $\tilde{q}$ is positive in a neighborhood of the singularity ($\tilde{q}$ can oscillate around zero; indeed, it is the possible oscillatory nature of the variables---i.e., $\tilde{q}$ need not be of single sign--- that causes potential problems). However, if $\tilde{q}\rightarrow 0$, (\ref{Omega_prime}) implies 
\begin{equation}
\tilde{\Omega}'=\tilde{H}\tilde{\Omega}(2-3\gamma)
\end{equation}
which implies a contradiction for $\gamma>\frac{2}{3}$ (i.e. $\tilde{\Omega} \not \rightarrow 0$ as $\tau\rightarrow-\infty$). Hence, as $\tau\rightarrow-\infty$, $\tilde{q}>0$, where $D\rightarrow\infty$ and 
\begin{equation}
\tilde{H}'=-\tilde{q}(1-\tilde{H}^2)
\end{equation}
and hence $\tilde{H}\rightarrow 1$ (assuming positive expansion) monotonically. [Note: this implies the existence of a monotonic function, and hence there are no periodic orbits  close to the singularity near the set $\tilde{H}=1$ ---all orbits approach $\tilde{H}=1$.] In addition, (\ref{constraint1}) gives $(\tilde{N}_1\tilde{N}_2\tilde{N}_3)\rightarrow 0$.

From Eqs. (\ref{Omega_prime}), (\ref{OmegaL_prime}), (\ref{Psi_prime}) we have as $\tau \rightarrow -\infty$
\begin{equation}
\frac{\tilde{\Phi}}{\tilde{\Psi}^2}\rightarrow 0,
\end{equation}
that is, the scalar field becomes effectively massless, and
\begin{equation}
\frac{\tilde{\Omega}}{\tilde{\Psi}^2}\rightarrow 0 \ \ \ \ \text{if} \ \ \ \ \gamma<2. 
\end{equation}
This follows directly from the evolution equation in the case of an exponential scalar field potential, $\bar \epsilon =\sqrt{\frac{3}{2}}k$, and follows for any physical potential for which $\bar \epsilon$ is bounded as $\tau \rightarrow-\infty$. Hence we obtain
\begin{equation}
\tilde{q}=2(\tilde{\Sigma}_+^2+\tilde{\Sigma}_-^2)+(3\Gamma-1)C^2
\end{equation}
where $\Gamma=2$, $C^2=\frac{a^2}{4}D^2\tilde{\Psi}^4$ if $\gamma<2$; $\Gamma=2$, $C^2=\frac{a^2}{4}D^2(\tilde{\Omega}+\tilde{\Psi}^2)^2$ if $\gamma=2$;  $\Gamma=\gamma$ and $C^2=\frac{a^2}{4}D^2\tilde{\Omega}^2$ if there is no scalar field, where $\tilde\Omega_{\lambda}\rightarrow C^2$ as $\tau\rightarrow -\infty$, and the Friedmann equation becomes
\begin{equation}
1-(\tilde{\Sigma}_+^2+\tilde{\Sigma}_-^2)-C^2
=\frac{1}{12}((\tilde{N}_1^2+\tilde{N}_2^2+\tilde{N}_3^2)-2(\tilde{N}_1\tilde{N}_2+\tilde N_1\tilde N_3+ \tilde N_2\tilde N_3 ))\geq 0
\end{equation}
From Eqs. (\ref{Omega_prime}), (\ref{Psi_prime}) we obtain
\begin{equation}\label{C}
(C^2)'=2C^2\tilde{H}[\tilde{q}-(3\Gamma-1)].
\end{equation}

We still have the possibility of $\tilde{\Sigma}_\pm$ or $\tilde{N}_i$ oscillating as the singularity is approached (as in the Mixmaster models).  A rigorous proof that oscillatory behaviour does not occur can be presented using the techniques of Rendall and Ringstrom \cite{Rendall1997CQG} and \cite{Ringstrom2000CQG} (using analytic approximations to the Brane-Einstein equations for $\tilde{\Sigma}_\pm$, $\tilde{N}_\alpha$; i.e., estimates for these quantities that hold uniformly in an open neighborhood of the initial singularity).  
We can then prove that $\tilde{\Sigma}_\pm \rightarrow 0$ as $\tau\rightarrow -\infty$, $\tilde\Omega_{\lambda}\rightarrow 1$, and we obtain the Brane Robertson-Walker source.

Alternatively, from equation (\ref{C}) we have that either $C^2\rightarrow 0$ or $\tilde q \rightarrow 3\Gamma-1$.  If $C=0$ then $\tilde{q}=2(\tilde{\Sigma}_+^2+\tilde{\Sigma}_-^2)$ and we can show that we obtain contradiction ($C=0$ implies $D=0$!). Hence $C^2\not=0$, and $ \tilde{q}\rightarrow (3\Gamma-1)$, 
so that 
\begin{equation}
\tilde{\Sigma}_+^2+\tilde{\Sigma}_-^2=0
\end{equation}
and it is a simple matter to show that $\tilde{N}_\alpha \rightarrow 0$ and again we obtain the Brane-Robertson Walker source.

It was shown earlier (for no perfect fluid and a scalar field with an exponential potential) that the Brane-Robertson-Walker solution is always an equilibrium point of the system and that local stability analysis shows that it is a local source.


\section{Concluding Remarks}

Assuming that $\tilde \Omega_{\cal U}\geq 0$, the future asymptotic behaviour of the Bianchi IX brane world containing a scalar field having an exponential potential is not significantly different than what is found in general relativity \cite{vandenHoogen:1999}.  We observe that for $0<k<\sqrt{2}$, the future asymptotic state is characterized by the power-law inflationary solution, and if $k>\sqrt{2}$ there no longer exists any equilibrium point representing an expanding model that is stable to the future.  We therefore conclude that if $k>\sqrt{2}$ then the Bianchi IX models must recollapse.  In \cite{vandenHoogen:1999} it was shown that if $k>\sqrt{2}$ then a collapsing massless scalar solution was a stable equilibrium point.  Here, in the Braneworld scenario, we have that this final end-point is the Brane-Robertson Walker solution.  However, we also observe that a typical model on its way to this final endpoint will asymptote towards a collapsing massless scalar field solution.  However, if $\tilde \Omega_{\cal U}< 0$, then a variety of new behaviours are possible including possible oscillating cosmologies \cite{CamposSopuerta:2001b}.

The past asymptotic behaviour of the Bianchi IX braneworld containing a scalar field (and ordinary matter) is significantly different than what is found in general relativity.  It is known that the Bianchi IX perfect fluid models approach a MixMaster attractor (Kasner saddles joined by Taub separatrices) towards the past and are known to have chaotic behaviour.  Here we observe that the Brane Robertson-Walker solution is a global source, and that there is no periodic behaviour near the initial singularity.

Due to the quadratic nature of the brane corrections to the energy momentum tensor, a rich variety of intermediate behaviour is possible in these two fluid models.  We note the existence of the dark radiation density ${\cal U}$  together with a scalar field is similar to previous analysis done in \cite{Billyard} on scaling models.  Here the equivalent equation of state is $p_{\cal U}=(\frac{4}{3}-1){\cal U}$.  It is known that the bifurcation value for these scaling models is $k^2=3\gamma$, which for $\gamma=4/3$ corresponds to a value $k^2=4$, a bifurcation value found in the analysis above.    

The intermediate behaviour of these multifluid models considered here is extremely complex, and is not discussed in detail here.  However, a more complete analysis (in which some of the intermediate behaviour is outlined) of the Bianchi type II models is currently under investigation \cite{vandenHoogenAbolghasem:2002,vandenHoogenIbanez:2002}.

\begin{acknowledgments}
Both AAC and RJvdH are supported by research grants through Natural Sciences and Engineering Research Council of Canada.  RJvdH wishes to acknowledge the support of the University Council on Research at St. Francis Xavier University.  RJvdH would like to thank the faculty and staff in the Department of Mathematics and Statistics at Dalhousie University during his stay while this work was being completed.
\end{acknowledgments}


\centerline{{\bf References}}


\section*{Appendix: Integrability conditions for ${\cal P}_{\mu\nu} =0, {\cal Q}_\mu
= 0$}

The complete set of four dimensional Brane equations for a general imperfect fluid energy momentum tensor is given in Maartens (see equations 26-29 and A1-A10 in \cite{Maartens:2000a}). If ${\cal P}_{\mu\nu} =0, {\cal Q}_\mu
= 0$, then in the case of a perfect fluid with an equation of state
$p = (\gamma-1)\rho$, we obtain the following equations
when the vorticity is zero ($\omega=0$):

\begin{subequations} 
\begin{eqnarray}
&&A_\mu = -\frac{(\gamma-1)}{\gamma} \frac{1}{\rho} \D_\mu\rho \,,\label{CA}
\\
&&\dot{\rho}+\gamma\Theta\rho=0\,,\label{C1}
\\ 
&&\dot{\cal U}+{\frac{4}{3}}\Theta{\cal U}=0\,,\label{C2}
\\
&& \frac{1}{3}\D_\mu{\cal U}+\frac{4}{3}{\cal U}A_\mu
=-\frac{1}{6} \kappa^4 \gamma \rho\D_\mu \rho\,,\label{C3}
\\
&&\dot{\Theta}+\frac{1}{3}\Theta^2+\sigma_{\mu\nu}
\sigma^{\mu\nu}-{\rm D}^\mu A_\mu+A_\mu
A^\mu+\frac{1}{2}\kappa^2(3\gamma-2) \rho - \Lambda \nonumber
\\
&&\qquad\qquad = -\frac{\kappa^2}{2\lambda}(3\gamma-1)\rho^2-
\frac{6}{\kappa^2\lambda}{\cal U}\,, \label{C4}
\\
&& \dot{\sigma}_{\langle \mu\nu \rangle }
+\frac{2}{3}\Theta\sigma_{\mu\nu}
+E_{\mu\nu}-\D_{\langle \mu}A_{\nu\rangle } +\sigma_{\alpha\langle
\mu}\sigma_{\nu\rangle }{}^\alpha- A_{\langle \mu}A_{\nu\rangle}
=0\,, \label{C5}
\\
&& \D^\nu\sigma_{\mu\nu}
-\frac{2}{3}\D_\mu\Theta  = 0 \,,\label{C9}
\\
 && \D^\nu E_{\mu\nu}
 -\frac{1}{3}\kappa^2\D_\mu\rho
 -[\sigma,H]_\mu = \frac{\kappa^2 \rho}{3\lambda} \D_\mu\rho +\frac{2}{\kappa^2\lambda}\D_\mu{\cal U}\,, \label{C12}
\end{eqnarray}

and the Gauss-Codazzi equations on the brane   
\begin{eqnarray}
&&R^\perp_{\langle
\mu\nu\rangle}+\frac{1}{3}\Theta\sigma_{\mu\nu}
-E_{\mu\nu} - \sigma_{\alpha\langle
\mu}\sigma_{\nu\rangle }{}^\alpha  
=0 \,, \label{C15}\\
&& R^\perp+
\frac{2}{3}\Theta^2-\sigma_{\mu\nu} \sigma^{\mu\nu}
-2\kappa^2\rho -2\Lambda = \frac{\kappa^2}{\lambda}\rho^2+
\frac{12}{\kappa^2\lambda}{\cal U}\,. \label{C16}
\end{eqnarray}
\end{subequations}

Using Eq.(\ref{CA}) , Eq. (\ref{C3}) becomes for $\gamma \neq 1$:
\begin{equation}
\label{calU}
\D_\mu {\cal U} = \left[ \frac{4(\gamma -1)}{\gamma} 
\frac{\cal U}{\rho} - \frac{\kappa^4 \gamma}{2} \rho  \right] 
\D_\mu \rho.  
\end{equation}
Taking the directional time derivative (i.e., a dot) of Eq. (\ref{calU}),
and using Eqs. (\ref{C2}) and (\ref{C3}), and 
using known relations for interchanging space and time
derivatives, e.g., 
\begin{equation}
h_\mu{}^\nu[\D_\nu f]^\cdot 
= h_\mu{}^\nu [\nabla_\nu + A_\nu]
\dot{f}
- \left[\frac{1}{3} \Theta h_\mu{}^\nu + 
\sigma_\mu{}^\nu  \right] \D_\nu f,
\end{equation} 
we obtain the integrability condition
\begin{eqnarray}
&&\left[\frac{4}{3}(3 \gamma -4) 
{\cal U} - \frac{\kappa^4 \gamma^2}{2} \rho^2  \right] \rho \D_\mu \Theta =\nonumber\\
&&\qquad\qquad  
\left[\frac{4}{3\gamma}(3\gamma -4) (\gamma-1)
{\cal U}+ \frac{\kappa^4\gamma}{6}(3\gamma-1)\rho^2\right]
\Theta \D_\mu \rho.
\label{condition} 
\end{eqnarray}
Taking the directional time derivative of (\ref{C9}), interchanging
space and time derivatives, and using Eqs. (\ref{condition}),
(\ref{C9}), (\ref{C12})
(and (\ref{C1}), (\ref{C2}) and (\ref{C4})), we obtain
\begin{equation}
\label{equalszero}
\frac{(\gamma-1)}{\gamma} [12 {\cal U} + \kappa^4
\rho^2] \frac{1}{\rho} \D_\mu\rho =0.
\end{equation}
Hence for $\gamma \neq 1$ and $[12 {\cal U} + \kappa^4
\rho^2] \neq 0$, we must have
$$ \D_\mu \rho = 0,$$
so that
$$ A_\mu = \D_\mu {\cal U} = \D_\mu \Theta = \D^\nu \sigma_{\mu \nu}
= \D_\mu \sigma^2 =0$$
and hence in general the brane is spatially homogeneous.

The special case
\begin{equation}
\label{empty}
\gamma = \frac{2}{3}, \quad 12 {\cal U} + \kappa^4 
\rho^2 = 0,
\end{equation}
in which the integrability conditions yield no constraints, corresponds
to the case in which there are no corrections to the general relativistic equations.

Alternatively, Eqs. (\ref{C1}) and (\ref{C2}) imply that
\begin{equation}
\label{doth}
{\cal U} = h \rho^{\frac{4}{3 \gamma}}; ~ \dot{h} = 0
\end{equation}
and so Eq. (\ref{calU}) becomes
\begin{equation}
\label{aDag}
\D_\mu h = \left(\frac{4}{3 \gamma} (3 \gamma -4) h
- \frac{\kappa^4 \gamma}{2} \rho^{2-\frac{4}{3 \gamma}} 
 \right) \frac{1}{\rho} \D_\mu \rho.
\end{equation}
Defining
$$ \Psi = \frac{1}{h} \rho^{\frac{2(3\gamma -2)}{3 \gamma}}  $$
and $a = \frac{2(3\gamma -4)}{(3 \gamma -2)}$ and 
$ b= \frac{3 \kappa^4 \gamma^2}{4(3 \gamma -2)}$,
we then obtain
(for $\gamma \neq \frac{4}{3}$)
\begin{equation}
\label{A4}
\D_\mu (\ln h) = \left\{-\frac{2}{\Psi} - \frac{b}{(1-a)[(1-a) -b \Psi]}
\right\} \D_\mu \Psi,
\end{equation}
which integrates to 
\begin{equation}
\label{integrate}
h = f \Psi^{-2} [1 + \alpha \Psi]^{\frac{2-3 \gamma}{4-3 \gamma}},
\end{equation}
where $\D_\mu f =0$ and $\alpha \equiv \frac{b}{1-a} = \frac{3 \kappa^4
\gamma^2}
{4(4-3 \gamma)}$.  This Eq. (\ref{integrate})
essentially yields the functional dependence for $\rho$. 
In the exceptional case $\gamma = 4/3$, $\rho$ is separable 
in local coordinates.  (From Eq. (\ref{aDag}) it can be seen that the 
case $\gamma = 2/3$ is also exceptional).

Eq. (\ref{condition}) then becomes
\begin{equation}
\label{exceptional}
\left[\frac{4}{3} (3 \gamma -4)h - 
\frac{\kappa^4 \gamma^2}{2} \rho^{2-\frac{4}{3\gamma}} 
\right]  \frac{1}{\Theta} \D_\mu \Theta = 
\left[ \frac{4(3 \gamma -4) (\gamma-1)}
{3\gamma} h
- \frac{\kappa^4 \gamma(\gamma -1/3)}{2} 
\rho^{2-\frac{4}{4 \gamma}} \right] \frac{1}{\rho} \D_\mu \rho.
\end{equation}
From Eqs. (\ref{integrate}) and (\ref{C2}), and using local coordinates
so that $h = h(x^\gamma)$, $f = f(t)$,
we obtain the functional forms for $\rho$ and $\Theta$, whence on
substitution
into 
Eq. (\ref{exceptional}) (for $\gamma \neq 4/3)$ we obtain a contradiction
if $\D_\mu \rho \neq 0$.

Finally, in the special case $\gamma =1$, in which 
$p =0$ and $A_\mu =0$, we obtain analogues of Eqs (\ref{calU})
and (\ref{condition}) (which do not follow for $A_\mu =0)$:
\begin{eqnarray}
 \D_\mu {\cal U} &=& -\frac{1}{2} \kappa^4 \rho \D_\mu \rho\\
 \left[ {\cal U} + \frac{3}{8} 
\kappa^4 \rho^2 \right] = \frac{1}{\Theta} \D_\mu \Theta 
&=&  \frac{1}{8} \kappa^4 \rho \D_\mu \rho.
\end{eqnarray}
In local coordinates we then find (for $\D_\mu \rho \neq 0)$
\begin{equation}
- \frac{2}{3}h = \left( 1 + \frac{3}{8} \kappa^4
\rho^{2/3} + \frac{3}{32}
\kappa^8 \rho^{4/3}  \right) 
\end{equation}
which implies that $\dot{\rho} =0$ (and hence $\Theta=0$)!


\begin{thebibliography}{10}

\bibitem{rubakov}
V. Rubakov and M. E. Shaposhnikov, Phys. Lett. B{\bf 159}, 22 (1985); J.
Polchinski, Phys. Rev. Lett. {\bf 75}, 4724 (1995); P. Horava and E. Witten, 
Nucl. Phys. B{\bf 460}, 506 (1996); N. Arkani-Hamed, S. Dimopoulos and 
G. Dvali, Phys. Lett. B{\bf 429}, 263 (1998); I. Antoniadis, N. Arkani-Hamed, 
S. Dimopoulos and G. Dvali, Phys. Lett. B{\bf 436}, 257 (1998); A. Lukas, 
B. A. Ovrut and D. Waldram, Phys. Rev. D{\bf 60}, 086001 (1999).

\bibitem{randall}
L. Randall and R. Sundrum, Phys. Rev. Lett. {\bf 83}, 3370 (1999); {\it ibid}
{\bf 83}, 4690 (1999); 
N. Arkani-Hamed, S. Dimopoulos, G. Dvali and N. Kaloper, Phys.
Rev. Lett. {\bf 84}, 586 (2000); A. Chamblin and G. W. Gibbons, Phys. Rev. 
Lett. {\bf 84}, 1090 (2000).

\bibitem{Maartens:2001b}
R. Maartens, V. Sahni, and T.D. Saini, Phys. Rev. D, {\bf 63}, 063509 (2001).

\bibitem{Maartens:2000a}
R. Maartens, Phys. Rev. D {\bf 62}, 084023 (2000).

\bibitem{Shiromizu:2000} T. Shiromizu, K.I. Maeda, and M. Sasaki,
Phys. Rev. D {\bf 62}, 024012 (2000);
M. Sasaki, T. Shiromizu, and K.I. Maeda, Phys. Rev. D {\bf 62},
024008 (2000).

\bibitem{CollinsHawking:1973}
C. B. Collins and S. W. Hawking, Ap. J., {\bf 180}, 317 (1973).

\bibitem{js86}
L.G. Jensen and J.A. Stein-Schabes, Phys. Rev. D {\bf 34}, 931 (1986).

\bibitem{rellis86}
T. Rothman and G.F.R. Ellis, Phys. Lett. {\bf B180}, 19 (1986).

\bibitem{cline}
J. M. Cline, C. Grojean and G. Servant, Phys. Rev. Lett. {\bf 83}, 4245 (1999);
R. N. Mohapatra, A. Perez-Lorenzana and  C. A. de S. Pires, Phys. Rev. D 
{\bf 62}, 105030 (2000); R. N. Mohapatra, A. Perez-Lorenzana and  
C. A. de S. Pires, Int. J. Mod. Phys. {\bf A}16, 1431 (2001); C. Csaki, M. Graesser, C. Kolda, and J. Terning,
Phys. Lett. {\bf B462}, 34 (1999); D. Ida, J. High Energy Phys.
{\bf 09}, 014 (2000); L.E. Mendes and A.R. Liddle, Phys. Rev. D {\bf
62}, 103511 (2000); 
N. Kaloper, Phys. Rev. D {\bf 60}, 123506 (1999);
T. Nihei, Phys. Lett. {\bf B465}, 81 (1999); H.B. Kim and H.D. Kim, Phys. Rev. D {\bf 61}, 064003
(2000); P. Kanti, I.I. Kogan, K.A. Olive, and M. Pospelov, Phys.
Lett. {\bf B468}, 31 (1999);  P. Kraus, JHEP {\bf 9912}, 011 (1999);
 S. Mukohyama, Phys. Lett. {\bf B473}, 241 (2000);
C. Csaki, M. Graesser, L. Randall, and J. Terning, Phys. Rev.
D {\bf 62}, 045015 (2000); C.
Barcelo and M. Visser, Phys. Lett. {\bf B482}, 183 (2000); J.
Lesgourgues, S. Pastor, M. Peloso, and L. Sorbo, hep-ph/0004086;
H. Stoica, S.-H. Henry Tye, and I. Wasserman, Phys. Lett. {\bf
B482}, 205 (2000).
S. Mukohyama, T. Shiromizu, and K. Maeda, Phys. Rev. D {\bf 62},
024028 (2000).

\bibitem{BinDefLan:2000}
P.~Bin$\acute{\mbox{e}}$truy, C.~Deffayet, and D.~Langlois, Nucl. Phys. {\bf
  B565},  269  (2000);
see also P.~Bin$\acute{\mbox{e}}$truy, C.~Deffayet, U.~Ellwanger, and D.~Langlois, Phys.
  Lett. B {\bf 477},  285  (2000),
$\acute{\mbox{E}}$.~$\acute{\mbox{E}}$.~Flanagan, S.H.~Henry.~Tye, and 
I.~Wasserman, Phys. Rev. D {\bf 62},  044039  (2000).

\bibitem{CamposSopuerta:2001a}
A. Campos and C.F. Sopuerta, Phys. Rev. D, {\bf 63}, 104012 (2001).

\bibitem{CamposSopuerta:2001b}
A. Campos and C.F. Sopuerta, Phys. Rev. D, {\bf 64}, 104011 (2001).

\bibitem{Coley:2002a}
A.A. Coley, Class. Quant. Grav., {\bf 19}, L45, (2002).

\bibitem{Coley:2002b}
A.A. Coley, Dynamics of Brane-world Cosmological Models, {Phys. Rev. D.}, {\bf 66}, 023512, (2002).

\bibitem{Santos:2001}
M.G. Santos, F. Vernizzi and P.G. Ferreira, Phys. Rev. D, {\bf 64}, 063506 (2001).

\bibitem{Maartens:2000b}
R. Maartens, D. Wands, B. A. Bassett, and I.P.C. Heard, Phys. Rev. D, {\bf 62}, 041301 (2000).

\bibitem{Copeland:2000}
E.J. Copeland, A.R. Liddle, and J.E. Lidsey, Phys. Rev. D, {\bf 64}, 023509 (2001).

\bibitem{Maeda:2002}
S. Mizumo, K.I. Maeda, and K. Yamamoto, Dynamics of scalar field in a brane world, {(hep-ph/0205292)}.

\bibitem{Dunsby:2002}
N. Goheer and P. Dunsby, Brane-world Dynamics of Inflationary Cosmologies with exponential Potentials, {(gr-qc/0204059)}.

\bibitem{Elliswainwright}
J. Wainwright and G.F.R. Ellis, {\it Dynamical Systems in Cosmology}, (Cambridge University Press, 1997).

\bibitem{La89a}  D. La and P.J. Steinhardt, Phys. Rev. Letts. {\bf 62}, 376 (1989).

 
\bibitem{Acetta90} P.J. Steinhardt and F.S. Accetta, Phys. Rev. Letts. {\bf 64}, 2740 (1990)

\bibitem{Salam84}  A. Salam and E. Sezgin, Phys. Letts. B {\bf 147}, 47 (1984)

\bibitem{Halliwell87} J.J. Halliwell, Phys. Letts. B {\bf 185}, 341 (1987)

\bibitem{Langlois:2001}
D. Langlois, R. Maartens, M. Sasaki, and D. Wands, Phys. Rev. D, {\bf 63}, 084009 (2001).

\bibitem{Gordon:2001}
C. Gordon and R. Maartens, Phys. Rev. D, {\bf 63}, 044022 (2001).

\bibitem{Rendall1997CQG}
A. Rendall, Class. Quantum Grav., {\bf 14}, 2341 (1997).

\bibitem{Ringstrom2000CQG}
H, Ringstrom  Class. Quantum Grav., {\bf 17}, 713 (2000).

\bibitem {vandenHoogen:1999}
R.J. van den Hoogen and I. Olasagasti, Phys. Rev. D, {\bf 59}, 107302 (1999).

\bibitem{Billyard}
A.P. Billyard, A.A. Coley, and R.J. van den Hoogen, Phys. Rev. D, {\bf 58}, 123501 (1998);
R.J. van den Hoogen, A.A. Coley, and D. Wands, Class. Quantum Grav., {\bf 16}, 1843 (1999).

\bibitem{vandenHoogenAbolghasem:2002}
R.J. van den Hoogen and H. Abolghasem, Bianchi II Braneworld Cosmologies ${\cal U}\geq 0$, preprint 2002.

\bibitem{vandenHoogenIbanez:2002}
R.J. van den Hoogen and J. Iba{\~n}ez, Bianchi II Braneworld Cosmologies ${\cal U}\geq 0$, preprint 2002.

 
\end{thebibliography}
\end{document}